\documentclass[twocolumn, aps,pre,superscriptaddress,nofootinbib]{revtex4-2}

\usepackage{pifont}

\usepackage{color, colortbl}
\definecolor{Gray}{gray}{0.2}

\usepackage{graphicx}
\usepackage[dvipsnames,svgnames]{xcolor}
\usepackage{bm,bbm}
\usepackage{braket}
\usepackage{amsthm,amsmath,amssymb}
\usepackage{enumerate}
\usepackage{booktabs,multirow}
\usepackage{mathtools,bbding}
\usepackage[percent]{overpic}
\usepackage{microtype}
\usepackage{array,adjustbox,afterpage}
\usepackage[T1]{fontenc}
\usepackage[utf8]{inputenc}
\usepackage{tikz}
\usetikzlibrary{calc}
\usepackage[english]{babel}
\usepackage{newtxtext,newtxmath}

\usepackage{lipsum}

\usepackage[linesnumbered,ruled]{algorithm2e}  

\makeatletter
\renewcommand{\fnum@figure}{FIG. \thefigure} 
\makeatother

\usepackage[bookmarks=false, colorlinks=true, linkcolor=blue, citecolor=purple, urlcolor=purple]{hyperref}
\usepackage{cleveref}
\usepackage[normalem]{ulem}

\usepackage{lipsum}
\usepackage{makecell}

\Crefname{subfigures}{figure}{figures}
\Crefname{subfigures}{Figure}{Figures}

\newcommand{\xdownarrow}[1]{{\left\downarrow\vbox to #1{}\right.\kern-\nulldelimiterspace}} 

\hypersetup{
    breaklinks=true,   
    colorlinks=true,   
    pdfusetitle=true,  
}

\begin{document}

\title{Witnessing superpositions of causal orders before the process is completed}
\author{Onur Pusuluk}
\email{onur.pusuluk@gmail.com}
\affiliation{Faculty of Engineering and Natural Sciences, Kadir Has University, 34083, Fatih, Istanbul, T\"{u}rkiye}
\author{Zafer Gedik}
\affiliation{Faculty of Engineering and Natural Sciences, Sabanci University, Tuzla, Istanbul 34956, T\"{u}rkiye}
\author{Vlatko Vedral}
\affiliation{Department of Physics, University of Oxford, Parks
Road, Oxford, OX1 3PU, UK}

\begin{abstract}

The questions we raise in this letter are as follows: What is the most general representation of a quantum state at a single point in time? Can we adapt the current formalisms to situations where the order of quantum operations is coherently or incoherently superposed? If so, what are the relations between the state at a given time and the uncertainty in the order of events before and after it? Establishing the relationship between two-state vector formalism and pseudo-density operators, we introduce the notion of a single-time pseudo-state. The tomographic construction of single-time pseudo-states is possible by ideal or weak measurements. We demonstrate that the eigenspectrum obtained from weak measurements enables us to discriminate between some coherent and incoherent superpositions of causal orders in pre- and post-selected systems before the process is completed. Finally, we discuss some possible experimental realizations in existing photonic setups.

\end{abstract}

\maketitle


\textit{Introduction.}--- In standard quantum mechanics, the state vector of a pre-selected system provides a probability distribution for the statistics of ideal measurements. It directly enters into the calculation of average values of physical observables. When the information on the pre-selected system is somehow less than the maximum, state vector formalism becomes insufficient. Consequently, we move on to the density matrix formalism to encode all measurement statistics by a single mathematical object. In this case, the density matrix, a statistical mixture of state vectors, has the minimum number of parameters to calculate the averages of observables. Algebraically, this matrix is a Hermitian, positive-semidefinite, and normalized operator that acts on the Hilbert space.

A single state vector is also inadequate to predict the measurement statistics of pre- and post-selected systems as it does not contain the restriction imposed by the post-selection measurement. In 1964, Aharonov-Bergmann-Lebowitz (ABL) showed in their seminal work~\cite{TSVF_1964} that ideal measurement of an observable \(\hat{C}\) at time \(t\) results in an outcome \(c_j\) with a probability given by
\begin{equation}\label{Eq_ABL}
p(c_j) = \frac{|\langle \phi |c_j\rangle \langle c_j|\psi \rangle|^2}{\sum_n |\langle \phi |c_n\rangle \langle c_n|\psi \rangle|^2} ,
\end{equation}
where the measurement outcomes \(\{c_j\}\) and corresponding post-measurement states \(\{|c_j\rangle\}\) constitute the eigendecomposition of the observable, \(|\psi\rangle\) is a forward evolving state determined by the results of pre-selection measurement performed at time \(t_1 < t\), and \(\langle \phi |\) is a backward evolving state determined by the results of post-selection measurement performed at time \(t_2 > t\). This formula led Aharonov and coworkers~\cite{TSVF_1991, TSVF_1995, TSVF_2008} to extend the state vector formalism to include backward evolving state \(\langle \phi|\). They describe the state of a pre- and post-selected system at time \(t\) by the so-called \textit{two-state vector}
\begin{equation}\label{Eq_TSVF}
\langle \phi| |\psi \rangle .
\end{equation}

In 1995, Reznik and Aharonov expressed the two-state in the form of a direct product of \(|\psi \rangle\) and \(\langle \phi|\)~\cite{TSVF_1995}, aiming to formulate the statistics of ideal and weak measurements~\cite{1988_WeakValue, 1990_WeakValue} in a unified manner. However, \(|\psi \rangle\langle \phi|\) is generally not a Hermitian, positive-semidefinite, or normalized operator. In a recent reformulation, Silva \textit{et al.} represented \eqref{Eq_TSVF} as \(_{t_2}\!\langle \phi| \otimes |\psi \rangle_{t_1}\)~\cite{2014_PRA_MixedTS}. To generalize the concept of two-state for incoherent mixtures \(\{p^r , \Psi^r\}\), they defined a \textit{density vector} such that \(\eta = \sum_r p^r \Psi^r \otimes \Psi^{r \dagger}\), where \(\Psi^r = \sum_{i, j} \alpha^r_{i j} \, _{t_2}\!\langle i| \otimes |j \rangle_{t_1}\) is a coherent superposition of two-states. This reformulation facilitated the demonstration that a subset of pre- and post-selected states corresponds to \textit{process matrices}~\cite{2012_ICO_NatCommun}, which are conventional mathematical objects used to characterize indefinite causal orders (ICOs) in the literature.

ICOs were initially proposed based on the notion that quantum gravity should be a probabilistic theory with a dynamic causal structure~\cite{2005_arXiv_Hardy, 2007_JPhysA_MathTheor_Hardy, 2011_JPhysConfSer_Hardy}. Hardy also suggested that computers operating without a definite causal structure could surpass conventional computers in performance~\cite{2009_Hardy}. The introduction of the quantum SWITCH has theoretically demonstrated in Refs.~\cite{2009_ICO_QSwitch, 2013_ICO_QSwitch} that such indefiniteness indeed offers computational advantages. The resource value of ICOs~\footnote{Note that the term \textit{indefinite causal structure} encompasses not only the superposition of different causal orders but also encompasses the superposition of various causal relations, including cause-effect and common cause relations~\cite{2017_NatCommun_Mixture_Cause-Effect_Common-Cause}.} has since been shown in various fields, including quantum computation~\cite{2012_PRA_Chiribella_qComp, 2012_PLA_qComp, 2015_SWITCH_Exp, 2017_PRA_qComp, 2021_PRX_qComp}, quantum communication~\cite{2018_PRL_EnhancedCommun, 2021_PRR_qCommun, 2021_NJP_qCommun}, quantum metrology~\cite{2020_PRL_qMet, 2021_PRA_qMet}, and quantum thermodynamics~\cite{2020_PRL_qThermo, 2020_PRA_qThermo, 2022_PRR_qThermo}. However, to efficiently harness this resource, it must first be identified. Current ICO witnesses in the literature~\cite{2015_CausalNonSep, 2017_ICO_Exp, 2018_ICO_Exp} can only detect this resource after the processes have been completed.

Our objective is to employ quantum state tomography to identify and characterize alternative mathematical objects that can encode measurement statistics for pre- and post-selected systems. By doing so, we aim to detect ICOs in these systems before the process is completed. One of us recently developed a similar approach in Ref.~\cite{2015_SRep_VV} that treats space and time equally in quantum mechanics. This approach assumes that the state of time-like separated systems can be reconstructed by a complete multi-time tomography. Moreover, the presence of negative eigenvalues in the resultant \textit{multi-time pseudo-density matrix}/\textit{operator} (PDO) is interpreted as evidence of temporal quantum correlations shared between the systems~\cite{2015_SRep_VV, 2018_PRA_VV, 2018_PRA_FNori}. Recently, mappings between PDOs, two-states, and process matrices have also been identified~\cite{2024_NJP_Oscar}. Here, we depart from these previous studies by considering complete single-time tomography of the system. Furthermore, we extend the tomographic construction to incorporate weak measurements~\cite{1988_WeakValue, 1990_WeakValue}. The resultant single-time PDO enables us to discriminate between some coherent and incoherent superpositions of causal orders in which the identical pre- and post-selection events interchange with a non-zero probability.


\textit{Incoherent superposition of causal orders}.--- Let us consider a pre- and post-selected ensemble of a two-level system for the sake of simplicity. Assume that two observables \(\hat{A}\) and \(\hat{B}\) are measured at times \(t_1\) and \(t_2\). The first (second) outcome is \(a\) (\(b\)) with unit probability and the measurement projects the state of the system onto the corresponding eigenstate of  \(\hat{A}\) (\(\hat{B}\)) represented by \(|a\rangle\) (\(|b\rangle\)). Alternatively, the system under consideration is pre-selected (post-selected) at time \(t_1\) (\(t_2\)) such that \(\hat{A} = a\) (\(\hat{B} = b\)).
\begin{figure*}[t] \centering
        \includegraphics[width=.8\textwidth]{Experiment.pdf}
        \caption{Two equivalent single-photon experiments for the description of incoherent mixture of causal orders by a single-time pseudo-density operator. The source of unpolarized photons is denoted by S, while D stands for the detector.  Pre- and post-selections are performed by vertical and diagonal polarizers represented by projectors \(|\psi\rangle \langle \psi |\) and \(|\phi\rangle \langle \phi |\). Pseudo-state tomography is based on the weak measurements of the photons between the polarizers. Time symmetry under the interchange of pre- and post-selection measurements are provided (a) by adding a mechanical coupling between the polarizers or (b) by joining the path with its reverse.}
        \label{Fig_Experiments}
\end{figure*}

The pseudo-state of the ensemble at an intermediate time \(t_1 < t < t_2\) can be tomographically constructed by
\begin{equation}\label{Eq_PDO_Wf}
\Gamma_\uparrow(t) = \frac{1}{2}\Big(\mathbb{I} + \sum_{j = 1}^3 \sigma^{j}_{w} \, \sigma^{j} \Big) = \frac{|\psi\rangle \langle \phi |}{\langle \phi | \psi \rangle} ,
\end{equation}
where \(\{\sigma^{j}\}\) are Pauli operators and the preceding weights are their weak values which are calculated by \(\sigma^{j}_w = \langle \phi | \sigma^{j} | \psi \rangle/\langle \phi| \psi \rangle \in \mathbb{C}\)~\cite{1988_WeakValue, 1990_WeakValue}. The backward- and forward-evolving quantum states used in this construction are defined by \(\langle \phi | \equiv \langle b | \hat{U}^\dagger(t,t_2)\) and \(|\psi\rangle \equiv \hat{U}(t_1,t) |a\rangle\). Here, \(\hat{U}\) is the unitary operator that describes the closed system dynamics between two measurement events. We assume an intermediate weak measurement in between these two ideal measurements that project the state on two non-orthogonal states \(|a\rangle\) and \(|b\rangle\). So, it is reasonable to expect that the two-state vector is not significantly disturbed at time \(t\). This is why we arrive at the normalized two-state in the direct product form in Eq.~(\ref{Eq_PDO_Wf}).

\(\Gamma_\uparrow(t)\) carries the whole information about the causal order between pre- and post-selection events, i.e., the reverse causal order results in the Hermitian conjugate of the two-state
\begin{equation}\label{Eq_PDO_Wb}
\Gamma_\downarrow(t) = \frac{|\phi\rangle \langle \psi |}{\langle \psi | \phi \rangle} = \Gamma_\uparrow^\dagger(t).
\end{equation}

Let us examine a scenario where two opposite causal orders, \(\Psi^{r=\uparrow} = \langle \phi| |\psi \rangle\) and \(\Psi^{r=\downarrow} = \langle \psi| |\phi \rangle\), coexist within an incoherent mixture. These causal orders are associated with the probabilities \(p^{r=\uparrow}\) and \(p^{r=\downarrow} = 1 - p^{r=\uparrow}\), respectively. We characterize this incoherent mixture using the density vector proposed in Ref.~\cite{2014_PRA_MixedTS} as:
\begin{equation}\label{Eq::Mixed2TimeState}
\begin{aligned}
  \eta_{\upharpoonleft\!\downharpoonright}(t) = &p^{r=\uparrow} |\phi\rangle_{t_2^\dagger} \otimes \, _{t_2}\!\langle \phi| \otimes |\psi \rangle_{t_1} \otimes \, _{t_1^\dagger}\!\langle \psi| \\
  &+ p^{r=\downarrow} |\psi\rangle_{t_2^\dagger} \otimes \, _{t_2}\!\langle \psi| \otimes |\phi \rangle_{t_1} \otimes \, _{t_1^\dagger}\!\langle \phi| .
\end{aligned}
\end{equation}
The weak value of an observable \(\hat{C}\) is then given by~\cite{2014_PRA_MixedTS} 
\begin{equation}\label{Eq::WV4Mixed2TimeState}
  \hat{C}^{\upharpoonleft\!\downharpoonright}_w = P(r=\uparrow\!|S) \hat{C}^\uparrow_w + P(r=\downarrow\!|S) \hat{C}^\downarrow_w ,
\end{equation}
where \(\hat{C}^\uparrow_w\) (\(\hat{C}^\downarrow_w\)) is the weak value of the same observable for the fixed causal order in which first \(|\psi\rangle\) (\(|\phi\rangle\)) is pre-selected at time \(t_1\) and then \(|\phi\rangle\) (\(|\psi\rangle\)) is post-selected at time \(t_2\). \(P(r=\uparrow\!|S)\) is the probability of the two-state being \(\Psi^{r=\uparrow}\) given the success of the post-selection:
\begin{equation}
  P(r=\uparrow\!|S) = \frac{p^{r=\uparrow} |\langle \phi|\psi \rangle|^2}{p^{r=\uparrow} |\langle \phi|\psi \rangle|^2 + p^{r=\downarrow} |\langle \psi|\phi \rangle|^2} = p^{r=\uparrow}.
\end{equation}
From now on, we will operate under the assumption that \(p^{r=\uparrow} = p^{r=\downarrow} = 1/2\). Consequently, the statistics of weak measurements described by \(\eta_{\upharpoonleft\!\downharpoonright}(t)\) can also be derived from the single-time pseudo-state below:
\begin{align}\label{Eq_PDO_W1}
\Gamma_{\upharpoonleft\!\downharpoonright}(t) &\equiv \frac{1}{2}\Big(\mathbb{I} + \sum_{j = 1}^3 (\sigma^{j})^{\upharpoonleft\!\downharpoonright}_w \, \sigma^{j} \Big) \, =  \frac{1}{2} \Gamma_\uparrow(t) + \frac{1}{2} \Gamma_\downarrow(t) ,
\end{align}
where \((\sigma^{j})^{\upharpoonleft\!\downharpoonright}_w\) is the \(\sigma^{j}\)'s weak value under the interchange of backward- and forward-evolving states with a probability of one half and equals to \((\sigma^{j}_w + \sigma^{j*}_w)/2 = \mathrm{Re}(\sigma^{j}_{w})\). \(\Gamma_{\upharpoonleft\!\downharpoonright}(t)\) is Hermitian and trace one by definition. Also, it has a negative eigenvalue unless \(\langle \phi |\psi \rangle \) is zero or one\footnote{\(\mathrm{tr}[\Gamma_{\upharpoonleft\!\downharpoonright}^2]\) equals to \(1 - 2 \det[\Gamma_{\upharpoonleft\!\downharpoonright}]\). Also, it is straightforward to show that \(\mathrm{tr}[\Gamma_{\upharpoonleft\!\downharpoonright}^2] = (1 + |\langle \phi | \psi \rangle|^{-2})/2\). So, the product of the eigenvalues of \(\Gamma_{\upharpoonleft\!\downharpoonright}\) should be negative when \(\langle \phi | \psi \rangle \neq 1\).}. So, the notion of positive-semidefinity can be relaxed not only for quantum states over time but also quantum states at a single-time. To our knowledge, this is something that has not been discussed yet in the literature.

If the intermediate measurement were ideal like pre- and post-selection measurements, even fixed causal order would result in a pseudo-state with a negative eigenvalue as shown in Appendix A. However, any mixed causal order cannot be distinguishable from fixed causal orders in this way.

The coherent superposition of causal orders can be experimentally realized in photonic setups using a quantum SWITCH~\cite{2017_ICO_Exp, 2018_ICO_Exp}. In these setups, if the initial state of the control qubit could be considered mixed, an incoherent mixture of causal orders could also be experimentally created. However, since the control qubit in these setups is typically represented by the path~\cite{2017_ICO_Exp} or polarization~\cite{2018_ICO_Exp} degree of freedom of a photon, preparing an initial mixed state can be complicated. Therefore, we propose alternative experimental setups for creating an incoherent mixture of causal orders.

Let us focus on photonic setups without loss of generality. Assume that our system of interest consist of unpolarized photons emitted from a single-photon source. Pre- and post-selection measurements are performed on the single photons through the vertical and diagonal polarizers placed between the light source and detector as shown in Fig.~\ref{Fig_Experiments}. By ignoring the free evolution of polarization in between the two ideal (projective) measurements and the weak measurement, we can write the backward- and forward-evolving states as
\begin{align}
\langle \phi | &= \cos(\theta) \langle V| + e^{-i \varphi} \sin(\theta) \langle H| \, , \label{Eq_Phi}\\
|\psi\rangle &= |H\rangle , \label{Eq_Psi}
\end{align}
where \(|H\rangle\) and \(|V\rangle\) represents the horizontal and vertical polarizations.

We will consider two different mechanisms that interchange the backward- and forward-evolving states with a probability of one half. First, we imagine that each polarizer can rotate between the configurations \(|\psi\rangle \langle \psi |\) and \(|\phi\rangle \langle \phi |\) as shown in Fig.~\ref{Fig_Experiments}-a. Random rotations of these polarizers can be synchronized by a mechanical coupling to ensure that they have always different configurations.

Alternatively, the symmetry under the interchange of backward- and forward-evolving states can be obtained by joining the path between the pre- and post-selection events with its reverse as shown in Fig.~\ref{Fig_Experiments}-b. In this case, four polarizers are placed between the photon source and detector in \(|\psi\rangle \langle \psi |\), \(|\phi\rangle \langle \phi |\), \(|\phi\rangle \langle \phi |\), and \(|\psi\rangle \langle \psi |\) configurations. The weak value is measured either between the first pair of polarizers or between the last pair of polarizers with an equal probability for each photon that hits the detector.

The two experiments proposed here result in identical pseudo-states for an ensemble of sufficiently large number of photons. It can be written in the basis of \(\{|V\rangle, |H\rangle\}\) as below
\begin{equation} \label{Eq_InCohMix}
\Gamma_{\upharpoonleft\!\downharpoonright}(\theta,\varphi) = \begin{pmatrix} 0 & \frac{1}{2} e^{-i \varphi} \cot(\theta) \\
\frac{1}{2} e^{i \varphi} \cot(\theta) & 1
\end{pmatrix} .
\end{equation}

Unlike \(\Gamma_\uparrow(\theta,\varphi)\) and \(\Gamma_\downarrow(\theta,\varphi)\), \(\Gamma_{\upharpoonleft\!\downharpoonright}(\theta,\varphi)\) has a negative eigenvalue as its determinant is given by \(- \cot^2(\theta)/4 \leq 0\). So, both experiments can be used to demonstrate the power of single-time PDOs to predict the measurement statistics of incoherent causal mixtures of the pre- and post-selection events.


\textit{Coherent superposition of causal orders}.--- A generic quantum superposition of two opposite causal orders \(\Psi^{r=\uparrow} = \langle \phi| |\psi \rangle\) and \(\Psi^{r=\downarrow} = \langle \psi| |\phi \rangle\) can be expressed as
\begin{equation}\label{Eq::Pure2TimeState}
  \psi_\uparrow \langle \phi| |\psi \rangle + \psi_\downarrow \langle \psi| |\phi \rangle ,
\end{equation}
where \(|\psi_\uparrow|^2 + |\psi_\downarrow|^2 = 1\). This superposition implies that the weak value of an observable \(\hat{C}\) at time \(t\) becomes~\cite{TSVF_1991}
\begin{equation}\label{Eq::WV4Pure2TimeState}
  \hat{C}^\updownarrow_w = q_{\uparrow} \hat{C}^\uparrow_w + q_{\downarrow} \hat{C}^\downarrow_w ,
\end{equation}
where 
\begin{subequations}
\begin{align}
q_{\uparrow} &= \frac{\psi_\uparrow \langle \phi | \psi \rangle}{\psi_\uparrow \langle \phi | \psi \rangle + \psi_\downarrow \langle \psi | \phi \rangle} ,
\\
q_{\downarrow} &= \frac{\psi_\downarrow \langle \psi | \phi \rangle}{\psi_\uparrow \langle \phi | \psi \rangle + \psi_\downarrow \langle \psi | \phi \rangle} .
\end{align}
\end{subequations}

\begin{figure}[t] \centering
        \includegraphics[width=.45\textwidth]{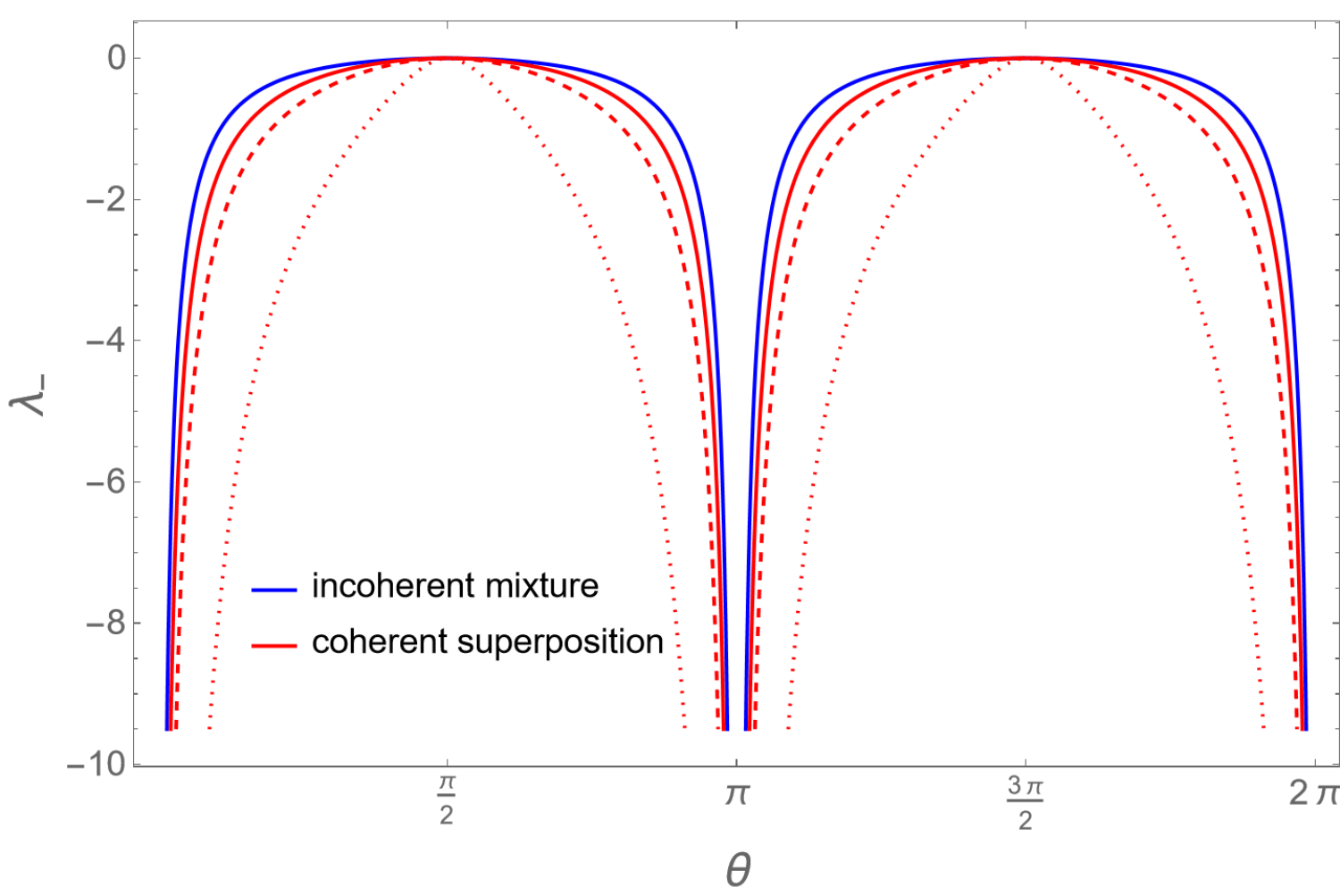}
        \caption{Negative eigenvalues of the single-time PDOs for incoherent (blue) and coherent (red) superpositions of the causal orders in which the pre- and post-selected states given in Eqs.~(\ref{Eq_Phi})~and~(\ref{Eq_Psi}) are interchanged with a probability of one half. The red solid, dashed, and dotted curves correspond to the cases where $\phi$ takes values of $\pi/4$, $\pi/3$, and $\pi/1.8$, respectively.}
        \label{Fig_Eigenvalues}
\end{figure}

Proceeding forward, we adopt the assumptions \(|\psi_{\uparrow}|^2 = |\psi_{\downarrow}|^2 = 1/2\) and \(q_{\uparrow} = q_{\downarrow}^\dagger\). Therefore, the statistical outcomes arising from weak measurements on Eq.~\eqref{Eq::Pure2TimeState} can be acquired through the utilization of the subsequent pseudo-state
\begin{eqnarray}\label{Eq_PDO_W2}
\begin{aligned}
\Gamma_\updownarrow(t) = \frac{1}{2}\Big(\mathbb{I} + \sum_{j = 1}^3 (\sigma^{j})^{\updownarrow}_w \, \sigma^{j} \Big)
\equiv q_{\uparrow} \Gamma_\uparrow(t) + q_{\downarrow} \Gamma_\downarrow(t) .
\end{aligned}
\end{eqnarray}
This representation of ~\eqref{Eq::Pure2TimeState} is Hermitian and trace one but not necessarily positive-semidefinite. Hence, not only incoherent mixture of two opposite causal orders but also their coherent superpositions can be witnessed by the eigenvalue spectrum of single-time PDOs. For example, the pseudo-state~(\ref{Eq_PDO_W2}) turns out to be
\begin{equation}\label{Eq_CohSup}
\Gamma_\updownarrow(\theta,\varphi) = \begin{pmatrix} 0 & q_\downarrow e^{-i \varphi} \cot(\theta) \\
q_\uparrow e^{i \varphi} \cot(\theta) & 0
\end{pmatrix} \, ,
\end{equation}
for the states \(|\phi\rangle\) and \(|\psi\rangle\) given in Eqs.~(\ref{Eq_Phi})~and~(\ref{Eq_Psi}). The eigenvalues of the pseudo-states \(\Gamma_{\upharpoonleft\!\downharpoonright}(\theta,\varphi)\) and \(\Gamma_\updownarrow(\theta,\varphi)\), as presented in Eqs.~(\ref{Eq_InCohMix})~and~(\ref{Eq_CohSup}), are given respectively by \(\{1/2\pm\csc(\theta)/2\}\) and \(\{1/2\pm\sqrt{1 + 4 |q_\uparrow|^2 \cot^2(\theta)}/2\}\). These two sets of eigenvalues coincides only when \(|q_\uparrow|^2 = 1/4\). Hence, coherent and incoherent superpositions of \(\Psi^{r=\uparrow}\) and \(\Psi^{r=\downarrow}\) become indistinguishable in this case. However, discerning between them is feasible when \(|q_\uparrow|^2 \neq 1/4\), for instance, when \(\psi_\uparrow = \psi_\downarrow = 1/\sqrt{2}\) and \(\langle \phi | \psi \rangle \in \mathbb{C}\). When the amplitudes \(\psi_\uparrow\) and \(\psi_\downarrow\) are equal, overlap \(\langle \phi | \psi \rangle\) becomes real only in the cases where \(\varphi = 0\) or \(\varphi = \pi\). In all other instances, the determinant of \(\Gamma_\updownarrow(\theta,\varphi)\), \(-\cot^2(\theta)\sec^2(\varphi)/4\), is smaller than the determinant of \(\Gamma_{\upharpoonleft\!\downharpoonright}(\theta,\varphi)\), indicating that the negative eigenvalue of \(\Gamma_\updownarrow(\theta,\varphi)\) (\(\lambda^\updownarrow_-\)) is less than the negative eigenvalue of \(\Gamma_{\upharpoonleft\!\downharpoonright}(\theta,\varphi)\) (\(\lambda^{\upharpoonleft\!\downharpoonright}_-\)), as shown in Fig.~\ref{Fig_Eigenvalues}. Hence, for the all values of the parameter \(\theta\), the magnitude of the single-time PDO's negative eigenvalue (\(|\lambda_-|\)) allows us to experimentally discriminate between the coherent and incoherent superpositions of the causal orders when we know the possible pre- and post-selected states.

The coherent superposition of two opposite causal orders can be generated by a quantum supermap called quantum SWITCH~\cite{2009_ICO_QSwitch, 2013_ICO_QSwitch}, which requires the presence of an ancillary two-level quantum system that acts as a control qubit. When the initial state of control qubit is changed from \(|0\rangle_c\) to \(|1\rangle_c\), the system of interest experiences two subsequent physical processes in a reversed order. Therefore, if the control qubit is prepared in a superposition state \((|0\rangle_c + |1\rangle_c)/\sqrt{2}\), then a coherent superposition of the opposite causal orders can be created.

Different optical implementations of quantum SWITCH were recently reported in Refs.~\cite{2017_ICO_Exp, 2018_ICO_Exp}. In this context, the photonic setups we proposed above can be extended to include quantum SWITCH. Alternatively, an entanglement swapping protocol, as proposed by Aharonov and coworkers~\cite{TSVF_1991, TSVF_2008} can also be utilized to generate a coherent superposition of opposite causal orders. This protocol similarly necessitates the presence of an ancillary two-level quantum system, akin to the SWITCH. Initially, the system and the ancilla are prepared in the entangled state \((|\Psi\rangle = |\psi\rangle |0\rangle_c + |\phi\rangle |1\rangle_c)/\sqrt{2}\) at time \(t_1\). Subsequently, the ancilla remains shielded from any interaction until the post-selection measurement at time \(t_2\), which projects the joint state onto \((|\Phi\rangle = |\phi\rangle |0\rangle_c + |\psi\rangle |1\rangle_c)/\sqrt{2}\). This projection effectively swaps the entanglement between the system and the ancilla to the two-state of the system starting from time \(t_2\) onward.

\textit{Thermal witnessing of indefinite causal orders.}--- Let us assume that the Hamiltonian of the system of interest is \(H_S = E_0 |0 \rangle \langle 0|+E_1 |1 \rangle \langle 1|\). Consider the following pre- and post-selected states which are purified by including some ancillary degrees of freedom \(A\)
\begin{eqnarray}
|\Psi(t_j)\rangle &= |\omega_j\rangle \equiv \sqrt{p_{0}} |0\rangle_S |0\rangle_{A} + i^{(j-1)} \sqrt{p_{1}} |1\rangle_S |1\rangle_{A} , \,
\end{eqnarray}
where \(p_{i} = e^{-\beta E_i}/(e^{-\beta E_0}+e^{-\beta E_1})\) and \(\beta = 1/ k_B T_S\). Notice that the reduced states of the system at times \(t_1\) and \(t_2\) are the same effective Gibbs state at inverse temperature \(\beta\).

\begin{figure*}[t] \centering
        \includegraphics[width=.9\textwidth]{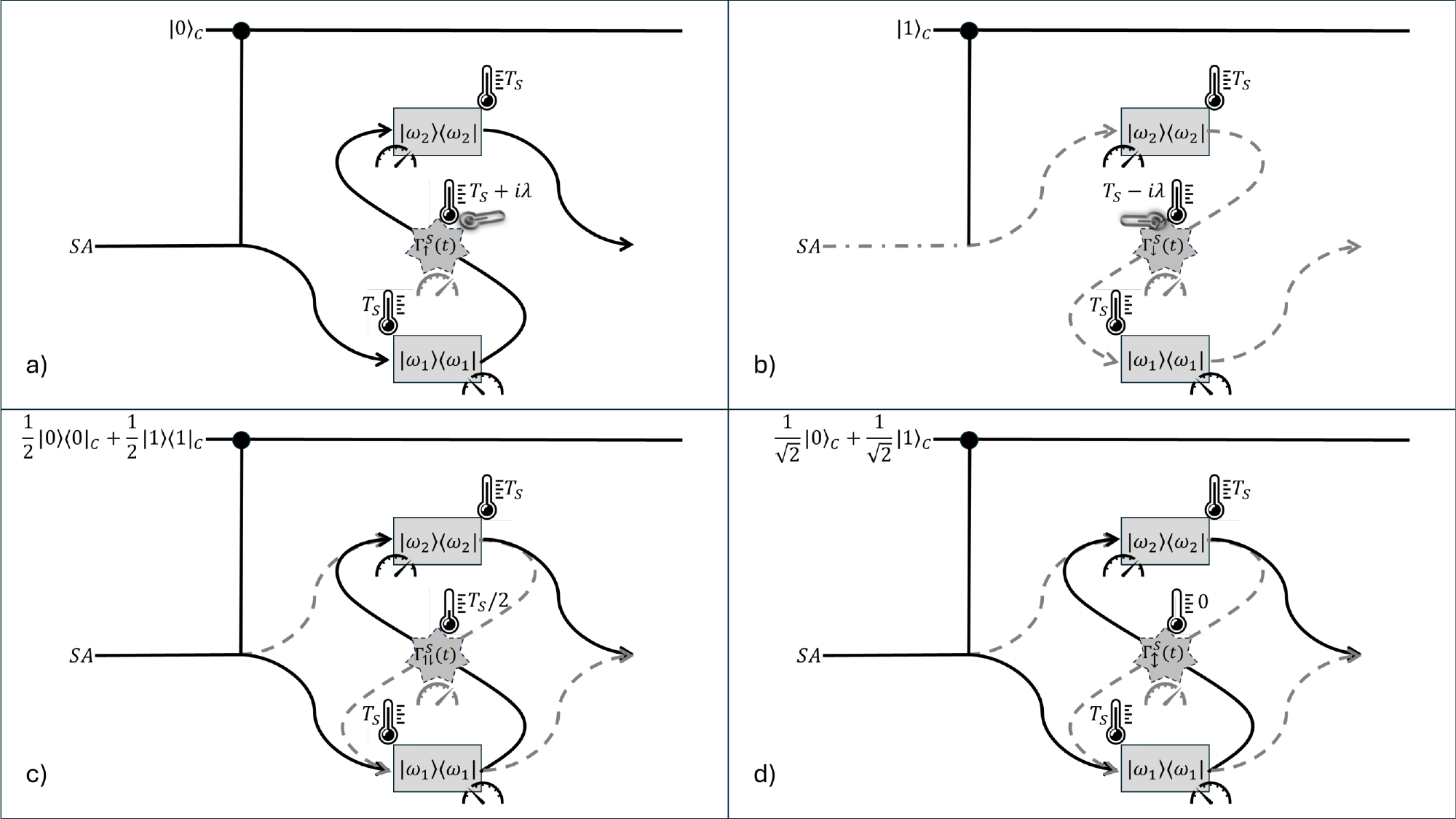}
        \caption{Thermal witnessing of indefinite causal orders. a) When the control qubit (\textit{C}) is in the \(|0\rangle\) state, the system of interest (\textit{S}) and an ancillary system (\textit{A}) are projected onto the entangled states of \(|\omega_1\rangle\) (\(|\omega_2\rangle\)) at time \(t_1\) (\(t_2\)). Through these projections, \textit{S} acquires an effective temperature \(T_S\).  If a weak measurement-based tomography is conducted at an intermediate time \(t\) to construct the pseudo-state of \textit{S}, the effective temperature gains an imaginary component. b) When \textit{C} is in the \(|1\rangle\) state, the order of the projection operations at times \(t_1\) and \(t_2\) changes. This change alters the imaginary component of the effective temperature at time \(t\). c) In the case where \textit{C} exists in \(|0\rangle\) and \(|1\rangle\) states with equal probabilities, an incoherent mixture of fixed causal orders in (a) and (b) is obtained. Consequently, the effective temperature at time \(t\) reduces to half of the temperature at times times \(t_1\) and \(t_2\) changes. d) If \textit{C} exists in a superposition of \(|0\rangle\) and \(|1\rangle\) states, it results in a quantum superposition of fixed causal orders in (a) and (b). Notably, in this scenario, the effective temperature at time \(t\) remains unaffected by the temperature observed at times \(t_1\) and \(t_2\), persistently holding at \(0\). In this example, the emergence of fixed causal orders with complex-valued temperatures and incoherent/coherent superpositions with real-valued temperatures can be attributed to the methodology utilized in weak measurement-based tomography. Notably, this approach focuses exclusively on a subsystem, \textit{S}, rather than the entire system. Expanding tomography to encompass the entangled \textit{SA} system reveals a distinct pattern: determinants of fixed causal order pseudo-states register as zero, while others exhibit negative determinants. This observation underscores the significance of considering the interplay between causal and spatial correlations.}
        \label{Fig_Thermal}
\end{figure*}

Consider the construction of a pseudo-state characterizing the system at an intermediate time instant \(t\) between \(t_1\) and \(t_2\) through weak measurement-based tomography. Can the effective temperature of this single-time pseudo-state offer insights into the order of operations carried out at \(t_1\) and \(t_2\)? Indeed, \(\Gamma_{\uparrow}(t) = \text{tr}_{A_1 A_2} [ |\omega_1 \rangle \langle \omega_2|/ \langle \omega_2|\omega_1 \rangle]\) and \(\Gamma_{\downarrow}(t) = \text{tr}_{A_1 A_2} [ |\omega_2 \rangle \langle \omega_1|/ \langle \omega_1|\omega_2 \rangle]\) represent effective Gibbs states characterized by complex inverse temperatures. Neglecting unitary time evolution between \(t_1\) and \(t_2\), we derive \(\beta_\uparrow (t) = \beta + i \, \pi / (2 \, \Delta E)\) and \(\beta_\downarrow (t) = \beta - i \, \pi / (2 \, \Delta E)\) where \(\Delta E = E_1 - E_0\). Thus, the imaginary component of the effective temperature at \(t\) can discern between two opposite causal orders, regardless of the specific values \(\beta\) at \(t_1\) and \(t_2\) (see Fig.~\ref{Fig_Thermal}-a~and~-b).

Could classical and quantum uncertainties in the causal order also manifest in the effective temperature of this single-time pseudo-state? While \(\Gamma_{\uparrow}(t)\) and \(\Gamma_{\downarrow}(t)\) posses complex temperatures, their incoherent and coherent superpositions, \(\Gamma_{\upharpoonleft\!\downharpoonright}(t)\) and \(\Gamma_\updownarrow(t)\), exhibit real inverse temperatures. Specifically, \(\beta_{\upharpoonleft\!\downharpoonright}(t) = 2 \, \beta\) and \(\beta_{\updownarrow}(t) = \infty\). This implies that in an incoherent mixture of two opposite causal orders, the temperature at an intermediate time will be measured to be half of the effective temperature \(T_S = 1/ k_B\,\beta\) being measured at both pre- and post-selection moments (Fig.~\ref{Fig_Thermal}-c). Furthermore, in the presence of a symmetric quantum superposition of causal orders, the intermediate temperature will always be zero regardless of the value of \(T_S\) (Fig.~\ref{Fig_Thermal}-d).


\textit{Open problems.}--- The conventional formalism employed in the literature to explore the quantum superposition of causal orders is grounded in process matrices~\cite{2012_ICO_NatCommun}. Two-time two-states and multi-time pseudo-states, as they can be mapped onto process matrices, are also capable of modeling indefinite causal orders~\cite{2017_NJP_ICO_TSVF, 2024_NJP_Oscar}. In this letter, we focused on weak measurements that can be performed within different causal structures at a single time. We demonstrated that single-time pseudo-states, which only describe the statistics of these measurements, can also provide information about the causal structure. Given their relatively compact size compared to multi-time two-states and multi-time pseudo-states, as well as their ability to identify the causal order prior to the completion of the process, these pseudo-states hold promise in effectively modeling more intricate causal structures.

The probability for each causal order is the same in both coherent and incoherent superpositions that we have considered so far. However, the asymmetric probabilities lead to a non-Hermitian representation of these superpositions, similar to the representation of fixed causal orders in equations~\eqref{Eq_PDO_Wf}~and~\eqref{Eq_PDO_Wb}. At this juncture, a natural question arises: Is it necessary to relax Hermiticity and/or normalization conditions as well to attain a unique mathematical construct capable of encoding all the weak measurement statistics of a general pre- and post-selected ensemble? Although this possibility is beyond the scope of the present letter, its investigation is a natural direction to pursue future work.

The presence of negative eigenvalues in a two-time pseudo-state serves as an indicator of quantum causal/temporal correlations between two time-like separated subsystems~\cite{2015_SRep_VV}. In this work, we extend this concept by demonstrating the definition of a single-time pseudo-state exhibiting a negative eigenvalue. This pseudo-state characterizes the intermediate-time weak measurement statistics of a pre- and post-selected system. Given the dependence of these statistics on pre- and post-selection measurements, it may be more precise to associate the eigenvalues of the single-time pseudo-state with quasi-probabilities rather than probabilities. Recently, Ban has provided an alternative perspective on our addressed scenario, revealing that due to the interference of opposite causal order, Kirkwood-Dirac quasi-probabilities can be measured~\cite{2021_PLA_WVsWithICOs}. This allows for access to weak values without necessitating weak measurements. Consequently, the single-time pseudo-states we construct possess the potential to encode more than just the statistics of weak measurements. We believe this constitutes an innovative facet of our work and warrants further verification.

Finally, it should be noted that the approach proposed herein is primarily applicable to ensembles generated in controlled laboratory settings, due to the necessity of post-selection. Consequently, it cannot be readily extended to describe single quantum systems exhibiting negative eigenvalues and complex entropies. Nevertheless, sequential weak measurements allow for the extraction of weak values without the need for post-selection~\cite{2019_Quantum_WVsWithoutWMs, 2021_Exp_WVsWithoutWMs}. Their potential applications in distinguishing different causal structures are also anticipated~\cite{2019_Quantum_WVsWithoutWMs}. Moreover, some recent discussions within the AdS/CFT correspondence framework suggest the potential existence of two-states in natural settings~\cite{2021_PRD_PseudoEntropy}. Some investigations have even claimed a link between the emergence of time in dS/CFT and the imaginary component of entropy in such two-states~\cite{2023_PRL_PseudoEntropy}. In light of these discussions, we posit that exploring the negativity of eigenvalues in both single- and multi-time pseudo-states could offer valuable insights into a more comprehensive understanding of temporal quantum processes.

\textit{Acknowledgments.}---O.P. and Z.G. express their gratitude to Bilimler K\"{o}y\"{u} in Fo\c{ç}a, where a part of this work was conducted. O.P. and V.V. also thank Marco Genovese and his team, especially Ivo Pietro Degiovanni, for their valuable contributions and insightful discussions during the early stages of this study.

\pagebreak
\appendix


\textit{Appendix A: Tomography with ideal measurements.}--- Single-time pseudo-state tomography of pre- and post-selected ensembles can be also performed by strengthening the measurement interaction at intermediate time \(t\). To this end, the weak values \(\sigma^{j}_w\) in Eq.~(\ref{Eq_PDO_Wf}) should be replaced with the ABL formula-based expectation values. This construction ends up with
\begin{equation}\label{Eq_PDO_ABL}
\mathrm{R}(t) = \frac{1}{2}\Big(\mathbb{I} + \sum_{j = 1}^3 \langle \sigma^{j} \rangle_{\mathrm{ABL}} \, \sigma^{j} \Big) \, ,
\end{equation}
where the expectation values of Pauli operators are calculated by the ABL formula given in Eq.~(\ref{Eq_ABL}), i.e. \(\langle \sigma^{j} \rangle_{\mathrm{ABL}} = p(\sigma^{j}=+1) - p(\sigma^{j}=-1)\).

The single-time pseudo-state \(\mathrm{R}(t)\) is connected to the conventional pseudo-states~\cite{2015_SRep_VV, 2018_PRA_VV, 2018_PRA_FNori} through the partial trace operation. The sequential measurements at times $t_1$, $t$, and $t_2$ can be used to construct a single-qubit PDO at three times
\begin{equation}\label{Eq_PDO_1Q3t}
\mathrm{R}_{ACB} = \frac{1}{2^3}\sum_{i, j, k = 0}^3 \langle \sigma_A^{i} \otimes \sigma_C^{j} \otimes \sigma_B^{k} \rangle \, \sigma_A^{i} \otimes \sigma_C^{j} \otimes \sigma_B^{k} \, ,
\end{equation}
with single-time marginals
\begin{align}
\mathrm{R}_{A} &= \text{tr}_{CB} [\mathrm{R}_{ABC}] = |a\rangle\langle a| \, , \\
\mathrm{R}_{C} &= \text{tr}_{AB} [\mathrm{R}_{ABC}] = \frac{1}{2}\Big(\mathbb{I} + \sum_{j = 1}^3 \langle \, \sigma^{j} \rangle_{\mathrm{ABL}} \, \sigma^{j} \Big) = \mathrm{R}(t) \, , \\
\mathrm{R}_{B} &= \text{tr}_{AC} [\mathrm{R}_{ABC}] = |b\rangle\langle b| \, .
\end{align}

By definition, \(\mathrm{R}(t)\) is a Hermitian and trace one operator. Likewise the three-time pseudo-state \(\mathrm{R}_{ABC}\) or its two-time marginals \(\mathrm{R}_{AB}\) and \(\mathrm{R}_{BC}\), it is also not necessarily positive-semidefinite. We can show this by taking \(|\psi\rangle = |0\rangle\) and \(|\phi\rangle = \cos(\theta) |0\rangle + e^{i \varphi} \sin(\theta) |1\rangle\) without any loss of generality. In this case, Eq.~(\ref{Eq_PDO_ABL}) turns into a single-time PDO
\begin{equation}
\mathrm{R}(\theta,\varphi) = \begin{pmatrix} 1 & \frac{1}{2} e^{-i \varphi} \sin(2 \theta) \\
\frac{1}{2} e^{i \varphi} \sin(2 \theta) & 0
\end{pmatrix} ,
\end{equation}
with a determinant that equals to \(- \cos^2(\theta) \sin^2(\theta) \leq 0\). This pseudo-state has a negative eigenvalue unless \(\langle \phi |\psi \rangle \) is zero or one.

What happens if the causal order between the pre- and post-selection events changes? Although \(\mathrm{R}_{CBA} \neq \mathrm{R}_{ABC}\), \(\mathrm{R}(t)\) remains the same. The probabilities described by the ABL formula is time-symmetrical, and so is the pseudo-state \(\mathrm{R}(t)\).

Now let us go one step further. We shall consider the incoherent mixture of two causal orders represented by the state \(1/2 \, \mathrm{R}_{ABC} + 1/2 \, \mathrm{R}_{CBA}\). Once again, the single-time marginal at time \(t\) gives the same \(\mathrm{R}(t)\). Hence, a single-time pseudo-state tomography cannot discriminate between the two opposite causal orders or their different mixtures if it is based on the ideal measurements.


\textit{Appendix B: Independence from the strength of measurement interaction.}--- Both single-time PDOs given in Eqs.~(\ref{Eq_PDO_W1})~and~(\ref{Eq_PDO_ABL}) are time-symmetric in the sense that they are invariant under the interchange of backward- and forward-evolving states. However, they don't match with each other in general. Is it possible to perform a single-time pseudo-state tomography which is independent of how strongly we measure the pre- and pos-selected ensemble at the intermediate time?

Note that the single-time pseudo-state tomographies that we have considered so far are based on the measurements of Pauli operators. As a matter of fact, this choice has a direct effect on the relation between \(\mathrm{R}(t)\) and \(\Gamma(t)\). For example, the former reduces to the latter when \(|\psi\rangle\) and \(|\phi\rangle\) are eigenvectors of two different Pauli operators.

To show the equivalence of \(\mathrm{R}(t) = \Gamma(t)\) under the restriction above, let us represent the eigenspace projectors of a Pauli operator \(\sigma^{j}\) by \(\Pi^{j\pm} = |c_{j\pm}\rangle \langle c_{j\pm}|\) such that \(\sigma^{j} |c_{j\pm}\rangle = \pm |c_{j\pm}\rangle\). This allows us to rewrite the ABL formula in terms of the weak values of these projectors as follows:
\begin{equation}
p(\sigma^j = \pm1) = \frac{|\Pi^{j\pm}_w|^2}{|\Pi^{j+}_w|^2+|\Pi^{j-}_w|^2} .
\end{equation}

Then, the ABL formula-based expectation values used in Eq.~(\ref{Eq_PDO_ABL}) become
\begin{equation} \label{Eq_ABL2W}
\langle \sigma^{j} \rangle_{\mathrm{ABL}} = \frac{|\Pi^{j+}_w|^2-|\Pi^{j-}_w|^2}{|\Pi^{j+}_w|^2+|\Pi^{j-}_w|^2} \equiv \frac{E_1}{E_2} ,
\end{equation}
where the denominator \(E_2\) equals to unity when \(|\psi\rangle\) and \(|\phi\rangle\) are eigenvectors of two different Pauli operators. On the other hand, the numerator in Eq.~(\ref{Eq_ABL2W}) always reads \begin{align}
E_1 &= |\Pi^{j+}_w|^2-|1-\Pi^{j+}_w|^2 , \nonumber \\
&= \mathrm{Re}^2(\Pi^{j+}_w) - (1 - \mathrm{Re}(\Pi^{j+}_w))^2 = 2 \, \mathrm{Re}(\Pi^{j+}_w) - 1 , \nonumber \\
&= \mathrm{Re}((2\,\Pi^{j+} - \mathbb{I})_w) = \mathrm{Re}(\sigma^j_w) = \langle \sigma^j \rangle_w ,
\end{align}
as \(\Pi^{j+}_w + \Pi^{j-}_w = \mathbb{I}_w = 1\). Consequently, we end up with \(\langle \sigma^{j} \rangle_{\mathrm{ABL}} = \langle \sigma^{j} \rangle_w\), which in turn implies that \(\mathrm{R}(t) = \Gamma(t)\).


\begin{thebibliography}{99}

\bibitem{TSVF_1964} Y. Aharonov, P. G. Bergmann and J. L. Lebowitz. Time Symmetry in the Quantum Process of Measurement, \textit{Phys. Rev.} \textbf{134}, B1410 (1964).

\bibitem{TSVF_1991} Y. Aharonov and L. Vaidman. Complete description of a quantum system at a given time. \textit{J. Phys. A: Math. Gen} \textbf{24}, 2315--2328 (1991).

\bibitem{TSVF_1995} R. Reznik and Y. Aharonov. Time-symmetric Formulation of Quantum Mechanics, \textit{Phys. Rev. A} \textbf{52}, 2538 (1995).

\bibitem{TSVF_2008} Y. Aharonov and L. Vaidman. The Two-State Vector Formalism: an Updated Review, in \textit{Time in Quantum Mechanics}, Vol 1. , LNP Springer Verlag (2008).
    
\bibitem{1988_WeakValue} Y. Aharonov, D.Z. Albert, and L. Vaidman. How the result of a measurement of a component of the spin of a spin-1/2 particle can turn out to be 100. \textit{Phys. Rev. Lett.} \textbf{60}, 1351 (1988).

\bibitem{1990_WeakValue} Y. Aharonov and L. Vaidman. Properties of a quantum system during the time interval between two measurements. \textit{Phys. Rev. A} \textbf{41}, 11 (1990).
    
\bibitem{2014_PRA_MixedTS} R. Silva, Y. Guryanova, N. Brunner, N. Linden, A. J. Short, and S. Popescu. Pre- and postselected quantum states: Density matrices, tomography, and Kraus operators. \textit{Phys. Rev. A} \textbf{89}, 012121 (2014).
    
\bibitem{2017_NJP_ICO_TSVF} R. Silva, Y. Guryanova, A. J. Short, P. Skrzypczyk, N. Brunner, and S. Popescu. Connecting processes with indefinite causal order and multi-time quantum states. \textit{New. J. Phys} \textbf{19}, 103022 (2017).
    
\bibitem{2012_ICO_NatCommun} O. Oreshkov, F. Costa, and \v{C}. Brukner. Quantum correlations with no causal order. \textit{Nat. Commun.} \textbf{3}, 1092 (2012).
    
\bibitem{2005_arXiv_Hardy} L. Hardy. Probability theories with dynamic causal structure: a new framework for quantum gravity. Preprint at arXiv:0509120 (2005).
    
\bibitem{2007_JPhysA_MathTheor_Hardy} L. Hardy. Towards quantum gravity: a framework for probabilistic theories with non-fixed causal structure. \textit{J. Phys. A} \textbf{40}, 3081 (2007).
    
\bibitem{2011_JPhysConfSer_Hardy} S. Markes and L. Hardy. Entropy for theories with indefinite causal structure. \textit{J. Phys.: Conf. Ser.}, \textbf{306}, 012043 (2011).
    
\bibitem{2009_Hardy} L. Hardy. Quantum gravity computers: On the theory of computation with indefinite causal structure. In: Quantum reality, relativistic causality, and closing the epistemic circle, \textbf{73}, 379--401, Springer, Dordrecht (2009).
    
\bibitem{2009_ICO_QSwitch} G. Chiribella, G.M. D'Ariano, P. Perinotti, B. Valiron. Beyond Quantum Computers. Preprint at arXiv:0912.0195 (2009).

\bibitem{2013_ICO_QSwitch} G. Chiribella, and G. M. D'Ariano, P. Perinotti, B. Valiron. Quantum computations without definite causal structure. \textit{Phys. Rev. A} \textbf{88}, 022318 (2013).

\bibitem{2017_NatCommun_Mixture_Cause-Effect_Common-Cause} J.-P. W. MacLean, K. Ried, R. W. Spekkens, K. J. Resch. Quantum-coherent mixtures of causal relations. \textit{Nat. Commun.} \textbf{8}, 15149 (2017).

\bibitem{2012_PRA_Chiribella_qComp} G. Chiribella. Perfect discrimination of no-signalling channels via quantum superposition of causal structures. \textit{Phys. Rev. A} \textbf{86}, 040301(R) (2012).
    
\bibitem{2012_PLA_qComp} T. Colnagh, G.M. D'Ariano, S. Facchini, and P. Perinotti. Quantum computation with programmable connections between gates. \textit{Phys. Lett. A} \textbf{376}, 2940--2943 (2012).
    
\bibitem{2015_SWITCH_Exp} L. Procopio, A. Moqanaki, M. Ara\'{u}jo, et al. Experimental superposition of orders of quantum gates. \textit{Nat. Commun.} \textbf{6}, 7913 (2015).
    
\bibitem{2017_PRA_qComp} M. Ara\'{u}jo, P.A. Gu\'{e}rin, and \"{A}. Baumeler. Quantum computation with indefinite causal structures. \textit{Phys. Rev. A} \textbf{96}, 052315 (2017).
    
\bibitem{2021_PRX_qComp} M.M. Taddei et al. Computational advantage from the quantum superposition of multiple temporal orders of photonic gates. \textit{PRX Quantum} \textbf{2}, 010320 (2021).
    
\bibitem{2018_PRL_EnhancedCommun} D. Ebler, S. Salek, and Giulio Chiribella. Enhanced communication with the assistance of indefinite causal order. \textit{Phys. Rev. Lett.} \textbf{120}, 120502 (2018).
    
\bibitem{2021_PRR_qCommun} G. Rubino et al. Experimental quantum communication enhancement by superposing trajectories. \textit{Phys. Rev. Research} \textbf{3}, 013093 (2021).
    
\bibitem{2021_NJP_qCommun} G. Chiribella et al. Indefinite causal order enables perfect quantum communication with zero capacity channels. \textit{New J. Phys.} \textbf{23}, 033039 (2021).
    
\bibitem{2020_PRL_qMet} X. Zhao, Y. Yang, and G. Chiribella. Quantum metrology with indefinite causal order. \textit{Phys. Rev. Lett.} \textbf{124}, 190503 (2020).
    
\bibitem{2021_PRA_qMet} F. Chapeau-Blondeau. Noisy quantum metrology with the assistance of indefinite causal order. \textit{Phys. Rev. A} \textbf{103}, 032615 (2021).
    
\bibitem{2020_PRL_qThermo} D. Felce and V. Vedral. Quantum refrigeration with indefinite causal order. \textit{Phys. Rev. Lett.} \textbf{125}, 070603 (2020).
    
\bibitem{2020_PRA_qThermo} T. Guha, M. Alimuddin, and P. Parashar. Thermodynamic advancement in the causally inseparable occurrence of thermal maps. \textit{Phys. Rev. A} \textbf{102}, 032215 (2020).
    
\bibitem{2022_PRR_qThermo} G. Rubino et al. Inferring work by quantum superposing forward and time-reversal evolutions. \textit{Phys. Rev. Research} \textbf{4}, 013208 (2022).

\bibitem{2015_CausalNonSep} M. Ara\'{u}jo, C. Branciard, F. Costa, A. Feix, C. Giarmatzi, and \v{C}. Brukner. Witnessing causal nonseparability. \textit{New J. Phys.} \textbf{17}, 102001 (2015).

\bibitem{2017_ICO_Exp} G. Rubino, L.A. Rozema, A. Feix, M. Ara\'{u}jo, J.M. Zeuner, L.M. Procopio, \v{C}. Brukner, P. Walther. Experimental verification of an indefinite causal order. \textit{Sci. Adv.} \textbf{3}, e1602589 (2017).
    
\bibitem{2018_ICO_Exp} K. Goswami et al. Indefinite causal order in a quantum switch. \textit{Phys. Rev. Lett.} \textbf{121}, 090503 (2018).

\bibitem{2015_SRep_VV} J. Fitzsimons, J. Jones, and V. Vedral. Quantum correlations which imply causation. \textit{Sci. Rep.} \textbf{5}, 18281 (2016).

\bibitem{2018_PRA_VV} Z. Zhao, R. Pisarczyk, J. Thompson, M. Gu, V. Vedral, and J.F. Fitzsimons. Geometry of quantum correlations in space-time. \textit{Phys. Rev. A} \textbf{98}, 052312 (2018).

\bibitem{2018_PRA_FNori} H.-Y. Ku, S.-L. Chen, N. Lambert, Y.-N. Chen, and F. Nori. Hierarchy in temporal quantum correlations. \textit{Phys. Rev. A} \textbf{98}, 022104 (2018).
    
\bibitem{2024_NJP_Oscar} X. Liu, Z. Jia, Y. Qiu, F. Li, O. Dahlsten. Unification of spatiotemporal quantum formalisms: mapping between process and pseudo-density matrices via multiple-time states. \textit{New. J. Phys.} \textbf{26}, 033008 (2024).
    
\bibitem{2021_PLA_WVsWithICOs} M. Ban. On sequential measurements with indefinite causal order. \textit{Phys. Lett. A} \textbf{403}, 127383 (2021).
    
\bibitem{2019_Quantum_WVsWithoutWMs} A. A. Abbott, R. Silva, J. Wechs, N. Brunner, C. Branciard. Anomalous weak values without post-selection. \textit{Quantum} \textbf{3}, 194 (2019).
    
\bibitem{2021_Exp_WVsWithoutWMs} M. Yang, Q. Li, Z.-H. Liu, Z.-Y. Hao, C.-L. Ren, J.-S. Xu, C.-F. Li, G.-C. Guo. Experimental observation of an anomalous weak value without post-selection. \textit{Photonics Res.} \textbf{8} (9), 1468--1474 (2020).
    
\bibitem{2021_PRD_PseudoEntropy} Y. Nakata, T. Takayanagi, Y. Taki, K. Tamaoka, and Z. Wei. New holographic generalization of entanglement entropy \textit{Phys. Rev. D} \textbf{103}, 026005 (2021). 

\bibitem{2023_PRL_PseudoEntropy} K. Doi, J. Harper, A. Mollabashi, T. Takayanagi, and Y. Taki. Pseudoentropy in dS/CFT and timelike entanglement entropy. \textit{Phys. Rev. Lett.} \textbf{130}, 031601 (2023). 

%
%
%
%
%
%
%
%

\end{thebibliography}
\end{document}